\documentclass[aps,twocolumn,prb,superscript,floatfix,superscriptaddress,showpacs,footinbib]{revtex4-2}

\usepackage{amssymb}

\usepackage[pdftex]{graphicx}

\usepackage{dcolumn}
\usepackage{bm}
\usepackage{amsmath}
\usepackage{array}
\usepackage{color}
\usepackage{float}
\usepackage{subfigure}
\usepackage{dsfont}
\usepackage{txfonts}
\usepackage{wasysym}
\usepackage{multirow}
\usepackage{sidecap}
\usepackage{xcolor,cancel}
\usepackage[normalem]{ulem}

\newcommand{\D}{\Delta}

\newcommand{\dn}{\downarrow}

\newcommand{\up}{\uparrow}

\newcommand{\meanval}[1]{{\langle{#1}\rangle}}

\usepackage{comment}

\newcommand{\ket}[1]{{|{#1}\rangle}}
\newcommand{\bra}[1]{{\langle{#1}|}}
\newcommand{\braket}[2]{{\langle{#1}|}{{#2}\rangle}}

\usepackage{hyperref}
\hypersetup{
    colorlinks,%
    citecolor=blue,%
    linkcolor=blue,%
    urlcolor=blue
}

\begin{document} 


\title{From ergodicity to Stark many-body localization in spin chains with single-ion anisotropy}

\author{M. G. Sousa}

\affiliation{Instituto de F\'isica, Universidade Federal de 
Uberl\^andia, Uberl\^andia, Minas Gerais 38400-902, Brazil.}

\author{Rafael F. P. Costa}
\affiliation{Instituto de F\'isica, Universidade Federal de 
Uberl\^andia, Uberl\^andia, Minas Gerais 38400-902, Brazil.}

\author{G. D. de Moraes Neto}
\email{gdmneto@zjnu.edu.cn}
\affiliation{Department of Physics, Zhejiang Normal University, Jinhua 321004, Zhejiang, P. R. China}

\author{E. Vernek}
\email{vernek@ufu.br}
\affiliation{Instituto de F\'isica, Universidade Federal de 
Uberl\^andia, Uberl\^andia, Minas Gerais 38400-902, Brazil.}
\affiliation{Department of Physics, Zhejiang Normal University, Jinhua 321004, Zhejiang, P. R. China}


\date{\today}

\begin{abstract}
The principles of ergodicity and thermalization constitute the foundation of statistical mechanics, positing that a many-body system progressively loses its local information as it evolves. Nevertheless, these principles can be disrupted when thermalization dynamics lead to the conservation of local information,  as observed in the phenomenon known as many-body localization.
Quantum spin chains provide a fundamental platform for exploring the dynamics of closed interacting quantum many-body systems. This study explores the dynamics of a spin chain with $S\geq 1/2$ within the Majumdar-Ghosh model, incorporating a non-uniform magnetic field and single-ion anisotropy. Through the use of exact numerical diagonalization, we unveil that a nearly constant-gradient magnetic field suppress thermalization, a phenomenon termed Stark many-body localization (SMBL), previously observed in $S=1/2$ chains. Furthermore, our findings reveal that the sole presence of single-ion anisotropy is sufficient to prevent thermalization in the system. Interestingly, when the magnitudes of the magnetic field and anisotropy are comparable, they compete, favoring delocalization. Despite the potential hindrance of SMBL by single-ion anisotropy in this scenario, it introduces an alternative mechanism for localization. Our interpretation, considering local energetic constraints and resonances between degenerate eigenstates, not only provides insights into SMBL but also opens avenues for future experimental investigations into the enriched phenomenology of disordered free localized $S\geq 1/2$ systems.

\end{abstract}
\maketitle
\section{Introduction}
In quantum mechanics, a given initial state of a closed interacting quantum system will evolve unitarily according its Hamiltonian $H$. A fundamental assumption in statistical physics is that a generic closed quantum many-body systems thermalize under its own dynamics~\cite{RevModPhys.83.863,DAlessio2016}. This implies that, under unitary time evolution, the reduced density matrix of any  generic initial state $\ket{\Psi(0)}$ tends to evolve towards the equilibrium Gibbs state within that subsystem.  Since any initial state can be spanned by eigenstates of the Hamiltonian, then the reduced matrix constructed with the eigenstates should also evolve toward equilibrium. This is the notion underlying the known eigenstate thermalization hypotesis (ETH)~\cite{PhysRevA.43.2046,PhysRevLett.80.1373,Rigol2008}.   
At a first glance, it seems that any out-of-equilibrium state evolves towards equilibrium. However, this is not always the case.
It was shown that disorder can induce localization in a variety of interacting systems~\cite{Alet2018,Chandran2014,Basko2006,Zangara2013,Bauer2013,Gornyi2005,Vasseur2016,Abanin2019,PhysRevB.75.155111,PhysRevB.77.064426,PhysRevB.80.115104,Vasquez2009,PhysRevLett.105.037001,Aleiner2010,PhysRevB.81.134202,PhysRevB.81.224429}, a phenomenon dubbed many-body-localization (MBL), as a generalization of the 
well known Anderson localization proposed by P. W. Anderson about fifty years ago~\cite{PhysRev.109.1492,Lagendijk2009,Kramer_1993}.

The intriguing phenomenon of ergodicity break in many-body localized systems has motivated great effort in the scientific community to understand the mechanisms that lead to MBL ~\cite{Nandkishore2015,Altman2015,Abanin2017,Smith2016}. Indeed, the phenomena of localization in interacting quantum systems is a puzzling problem still under study~\cite{PhysRevE.102.062144,Huang2022,PhysRevB.93.134201,PhysRevB.94.184202,Abanin2021,Panda2020,PhysRevLett.124.186601,PhysRevLett.125.156601,PhysRevLett.124.243601,PhysRevB.103.024203}. 
From a theoretical point of view, it is well established that the breakdown of the ETH in disorder induced many-body localized systems can be captured by the level-spacing statistics~\cite{DAlessio2016,Rigol2008,PhysRevLett.121.038901,Huang2022} or entanglement properties of their eigenstates~\cite{PhysRevB.77.064426,PhysRevLett.109.017202,Bauer2013}. The ETH usually understood in two classes: strong sense, in which all states thermalizes and weak sense, where almost all states thermalizes.  The main features of disorder-induced MBL phases have been observed in various experimental platforms~\cite{smith2016many,xu2018emulating,choi2016exploring,luschen2017observation,gong2021experimental}.  In spite of the complexities in the MBL phenomena, theoretical and experimental results suggest that, at least in  one dimension,  strong disorder induce the emergence of nearly conserved local quantities, leading to integrability~\cite{PhysRevLett.111.127201,PhysRevB.90.174202,PhysRevLett.116.010404,Imbrie2016,PhysRevB.94.144208}.
%

Recently, Pollman {\it et al}. showed that localization can occur in interacting systems even in the absence of disorder~\cite{Schulz2019}. The key ingredient here is the presence of a nearly uniform gradient potential in an interacting system  and this is closely tied to the single-particle localization process known as Wannier-Stark localization~\cite{vanNieuwenburg2019}.This phenomenon was called Stark Many-Body Localization (SMBL), and it shares similarities with the traditional MBL, such as level statistics, but differs in aspects such as a strong dependence on the initial conditions, as shown in~\cite{Doggen2021} for a spin $1/2$ Heisenberg chain. The mechanism leading to ETH violation in SMBL was believed to be Hilbert space fragmentation \cite{Doggen2021,PhysRevB.101.174204,PhysRevX.10.011047,PhysRevB.102.054206, Herviou2021}, in which the Hilbert space fragments into many disconnected subspaces, preventing themalization. However, this argument have been questioned~\cite{PhysRevB.105.L140201}.   Striking
 experiments in quantum simulators were performed to demonstrate Stark MBL~\cite{Morong2021,guo2021stark} and confirm that localization can arise in disorder-free systems.
 
In a recent study, one of us  have extended the results of Ref.~\cite{Doggen2021} by including exchange interaction up to second nearest neighbors~\cite{Vernek2022}. It was  showed that the SMBL phenomenon is robust for arbitrary ration $J_2/J_1$, a feature  also found in spinless fermions with long-range interactions~\cite{jiang2023stark}. SMBL has been also investigated in the context of bosonic lattices~\cite{yao2020many,taylor2020experimental} and it’s worth noting that other mechanisms can give rise to MBL, such as quasi-periodic potentials\cite{zhang2018universal,singh2021local} and periodic driving\cite{bairey2017driving,yousefjani2023floquet}.

Nonetheless, it is remarkable that all existing studies on spin lattices have exclusively focused on SMBL in spin  $1/2$ systems, which does not allow interactions between the systems' elementary excitations, \textit{kinks} or \textit{spinors}. Interactions arise naturally, in systems with spin $S\geq 1$, for instance, magnon-magnon interactions are present in spin$-1$ FM chains, giving rise to a nonlinear term $D\sum_{j=1}^L \left(S_j^{z}\right)^2$, where D represents the uniaxial anisotropy strength. Interestingly, weak ergodicity breaking in the form of many-body quantum scars appears in the spin$-1$ XY model~\cite{schecter2019weak}.
We should mention  that investigations on magnon-bound states has been conducted examining the impact of a weak anisotropy ($D$ 
 much smaller than the exchange interaction $J$) on thermalization time scales of spin$-1$ systems~\cite{wu2022exact,sharma2022multimagnon}.
Notably,  tunable single-ion anisotropy have been realized experimentally in trapped ions~\cite{senko2015realization}, ultra-cold atoms~\cite{chung2021tunable}, and compounds such as NiNb$_2$O$_6$~\cite{chauhan2020tunable} and [Ni(HF$_2$)($3-$Clpyradine)$_4$]BF$_4$~\cite{pajerowski2022high}.

In this paper, we investigate localization of quantum states in a  Majumdar-Ghosh model~\cite{10.1063/1.1664978} in the presence of a non-uniform magnetic field, and incorporating an additional term to address uniaxial single-ion anisotropy, which naturally arises in magnetic models with $S \geq 1/2$~\cite{Li2019,Liu2014,Nardelli2020}. By employing numerically exact diagonalization, we calculate the temporal evolution of complementary quantifiers, namely imbalance, entanglement entropy, and participation entropy. These measures are commonly used to characterize the ergodic-to-MBL phase diagram in quantum many-body systems. Our results reveal that for some initial product states: (i) Similar to the case of $S=1/2$ chain, MBL induced by finite $h_0$ is also capable of localizing system of spin $S>1/2$; (ii) For $S \geq 1 $, uniaxial  anisotropy alone localizes the system; (iii) If the system is localized for a particular value of $h_0$ the presence both $D \sim h_0$ suppresses localization and (iv) for $D,h_0 \gg J$ the system localizes. These results suggests that  the localization phenomena for  both  $D$ and $h_0$ is of the same nature. Indeed, they seem to result from a suppression of the spin dynamics by local energy constraint. While our numerically exact approach is limited by the size of the system, our results  suggest a different way of inducing localization in a quantum many-body system. Moreover, our results shed light on the nature of SMBL currently under intense investigations and paves the way for future experimental investigations  of these phenomena.  

The rest of this paper is organized as follow: In Sec.~\ref{model} we introduce the model and numerical methods, in Sec.~\ref{results} we present our numerical results and discussions. Finally, our work is summarized in Sec.~\ref{conclusions}

\section{Model and methods}\label{model}
For concreteness, we consider a spin chain modeled by an extended  version of  Majundar-Ghosh Hamiltonian~\cite{Bonechi1992}, which describes a physical system composed of a one-dimensional chain of interacting spins. Explicitly, the Hamiltonian for $L$ spins can be written as
\begin{eqnarray}\label{H_J1J2}
H=J_1\sum_{j=1}^{L-1} {\bf S}_j \cdot {\bf S}_{j+1}+ J_2\sum_{j=1}^{L-2} {\bf S}_j \cdot {\bf S}_{j+2}+\sum_{j=1}^L \left(h_j+DS_j^{z}\right)S_j^{z},
\end{eqnarray}
where the first two terms of our model Hamiltonian represent the exchange interaction between neighboring spins, capturing the mutual influence of these adjacent entities. $J_1$ and $J_2$ define the coupling strength between nearest and next-nearest neighbors, respectively. These are the two terms that corresponds to  what is known as Majumdar-Ghosh Hamiltonian~\cite{10.1063/1.1664978}. In the last summation of Eq.~\eqref{H_J1J2}, $h_jS_j^{z}$ (with $h_j=jh_0+\gamma j^2/L^2$) introduces the effect of a non-uniform magnetic field along the $z$-direction, allowing the study of Stark-many-body localization. Lastly, $D$ accounts for a single-ion anisotropy, which takes into account the directional preference of spins along the z-axis. This particular form of the Hamiltonian~\eqref{H_J1J2}, provides a theoretical framework for investigating fundamental properties of the system for spin $S=1/2$ and $S\geq 1$ as well. We should mention that the effect of $D$ can only be observed in the dynamics of system of spins $S>1/2$. In the special case of $S=1/2$, the effect of $D$ is just to shift the spectrum of the Hamiltonian by an amount $DL/4$.
 
%

%
For convenience, following Burssil et al~\cite{Bursill_1995} and our previous work~\cite{Vernek2022}, we  parameterize the exchange couplings $J_1$ and $J_2$ as $(J_1, J_2) = (J_0 \cos \theta, J_0 \sin \theta)$, with $0 \leq \theta < 2\pi$. The ground state properties for $h_0 = \gamma = 0$ in Eq.~\eqref{H_J1J2} have received great attention over the past decades~\cite{PhysRevB.66.024406,Vernek2020} However, our main focus here are thermalization processes, which potentially involves all eigenstates of the system consistent with the particular symmetry of the initial state. The absence of translation-invariance due to a finite field gradient naturally lead us to adopt open boundary conditions. This choice allows one to explore static and dynamical properties of the system in its entirety, with the price of finite sizes effects. Here we seek to understand how the system thermalizes, i.e., how it evolves from a given initial non-equilibrium state towards a thermal equilibrium state. More specifically, we will cover initial states of the form 
$\ket{\Psi_0}=~\mid \cdots m,m,m,\bar m,\bar m ,\bar m,\cdots\rangle$ in which $m=2S^z_j$ and $\bar m=-m$. This particular type of states correspond to a state with a single domains wall, but states with multiple domain walls will also be considered. This type of initial state has already been analyzed in other studies where it was possible to unveil several interesting characteristics of Stark many-body localization phenomena~\cite{Doggen2021,Vernek2022}. Here we will explore mainly the case of $S > 1/2$  by considering the effects of the single-ion anisotropy $D$ and   magnetic fields gradients $h_0$ in order to gain a better understanding of both local and global effects over the collective behavior of the spin dynamics. 

\subsection{Time-evolution analysis}
\label{Sec1A}
To study the phenomena of localization and thermalization in our closed quantum system, we  adopt the Schrödinger representation and perform time evolution of a given initial state  $\ket{\Psi_0}\equiv \ket{\Psi(t=0)}$ at $t=0$ as $\ket{\Psi(t)}=U(t)\ket{\Psi_0}$ where $U(t)=\exp{\left(-iHt/\hbar\right)}$ is the time evolution operator. Having the time-evolved quantum state $\ket{\Psi(t)}$, we are able to calculate the relevant physical quantity that witnesses localization/thermalization phenomenon, such as imbalance, participation and entanglement entropies. Since they are in general individually inconclusive  they are used complimentaryly.
\paragraph{Imbalance.--}
Imbalance,  denoted as ${\cal I}(t)$, is a key measure in investigating the dynamics of quantum systems, particularly in the context of Heisenberg models~\cite{Doggen2021,Morong2021,Schreiber2015,Schulz2019}. It provides insights on whether the system retains or loses information about local magnetization as it evolves in time. 
We define the imbalance for a chain of $L$ spins $S$ as
\begin{eqnarray}\label{imbalance}
{\cal I}(t)=\frac{1}{LS^2}\sum_{j=1}^L \langle S_j^z(t)\rangle\langle S_j^z(0)\rangle,
\end{eqnarray} 
where, $\meanval{S^z_j(t)}=\langle \Psi(t) \mid S_j^z \mid \Psi(t) \rangle$ represents the expectation value of  $S^z$ of the $j$-th spin of the chain within the evolving quantum state $\mid\Psi(t)\rangle$. As such, we clearly have ${\cal I}(0)=1$ if local spins are fully polarized, along either the positive or negative $S^z$ projections. Moreover, in the regime of strong localization, ${\cal I}(t)$ remains close to  its value at t=0, indicating that information about local magnetization is preserved during the evolution. On the other hand, in the thermal regime, ${\cal I}(t)$ tends to some lower value at long time scales. The precise asymptotic value of the imbalance in the thermal regime depends on the initial state. For instance, let us assume an initial product state $\ket{\Psi_0}$ as defined above for a chain of $L$ spins, where $N_{\up}$ and $N_{\dn}$ represent the number of spins with maximum and minimum $S^z$, respectively. Thus, the total spin projection along $z$-axis is given by $S^z_{\rm tot}=(N_{\up}-N_{\dn})S$. Since in the thermal state, this quantity will be uniformly distributed along the chain, each site will exhibit a magnetization $S^z_{\rm tot}/L$, if total $S^z$ is conserved. With this, we can  easily show that the thermal value for the imbalance is given by
\begin{eqnarray}
{\cal I}_{\rm thermal}=\left( \frac{S^z_{\rm tot}}{LS}\right)^2.  
\end{eqnarray}
In particular, for an initial state in the sector $S^z_{\rm tot}=0$ we obtain ${\cal I}_{\rm thermal}=0$. We should remark that, as defined in Eq.~\eqref{imbalance}, the imbalance is not capable of detecting  dynamics  of spins that occurs only on the $xy$-plane. In this case ${\cal I}(t)=0$, regardless the dynamics of the initial state. 

%
\paragraph{Entanglement entropy.--}
Another important quantity to monitor the localization of quantum states is the well known entanglement entropy $S_\ell(t)$ defined as

\begin{eqnarray}
S_\ell(t)=-\frac{1}{\ln 2}{\rm Tr}\left[\rho^{\ell}_A(t) \ln\rho^{\ell}_A(t)\right],
\end{eqnarray}
where $\rho^{\ell}_A(t)Tr_B[\rho^\ell_B]$ is the reduced density matrix of a subsystem $A$. Here, given the 1D nature of a chain, $\ell$ simply represents the site that defines the interface between two portions of the chain which we call subsystem $A$ and $B$. 

While the imbalance gives qualitative local data that is good for describing the dynamics of individual states, the use of entanglement entropy for many-body systems allows for the investigation of their collective and emergent properties, providing insights into the nature of quantum correlations and how information distributes throughout the entire system~\cite{Eisert2010}.


\paragraph{Participation entropy.--}
Participation entropy (PE) has gained attention in  studies of  dynamics of many-body systems. Here we follow Ref.~\cite{vanNieuwenburg2019} and defined this quantity as
\begin{eqnarray}\label{IPE}
S_2 = -\ln\left(\sum_{mn}\lvert c_n \rvert^2 \lvert c_m \rvert^2\delta_{E_n,Em}\right), 
\end{eqnarray}
where $c_n=\braket{n}{\Psi_0}$  is  the projection of the initial state $\ket{\Psi_0}$ onto the eigenbasis  $\{\ket{n}\}$. 
For practical purpose,  to calculate  $S_2$ we first define the quantity $f(t)=\bra{n}\exp(-iHt/\hbar)\ket{\Psi_0}$, which is projection of time evolved initial state onto the eigenstate $\ket{n}$. Upon Fourier transforming $f(t)$ to the energy domain, we obtain we obtain $F(\omega)$. The inverse participation ratio (IPR) can be obtained as
\begin{eqnarray}
IPR=\int F^2(\omega) d\omega=\sum_{mn}\lvert c_n \rvert^2 \lvert c_m \rvert^2\delta_{E_n,Em}.
\end{eqnarray}
With this, the participation entropy \eqref{IPE} is just $S_2=-\ln(IPR)$. For non degenerate case, this quantity coincides with the quantity $S_q$ for $q=2$ defined in Ref.~\cite{Mac2019}.
Defined as such, $S_2$ provides a measures of how the projections of the initial state are distributed among the eigenstates of the Hamiltonian.  In the absence of symmetries, for a delocalized state,  $S_2 \sim \ln ({\cal N})$, where ${\cal N}$ the dimension of the Hilbert space. This increases logarithmicaly  with the system size. On the other hand, in a strongly localized state, only a limited subset of coefficients contributes significantly to the sum, resulting in a constant PE $S_2$. In other words,  the state  is fully spanned on a restricted region of the Hilbert space, indicating the presence of localization. 

\section{Numerical results}\label{results}

To show our numerical results for the time evolution of the initial many-body state, we first set $J_0=1$ as energy unit and assume $\hbar=1$, so that time has units of inverse energy. Since the system is closed, exact numerical unitary evolution can be performed for small chains. To this end, here we use the python package Quspin \cite{10.21468/SciPostPhys.2.1.003,10.21468/SciPostPhys.7.2.020} that allows us to perform time evolution and calculation all the physical quantities we need. Since the Hilbert space increase as $(2S+1)^L$, where $L$ is the length of the chain, for a spin-1/2, for instance we can handle chains up to $L=22$ spins or so, depending on the symmetry  of the relevant sector of the Hilbert space. 
\begin{figure}[t!]
	\centering
	\subfigure{\includegraphics[clip,width=3.40in]{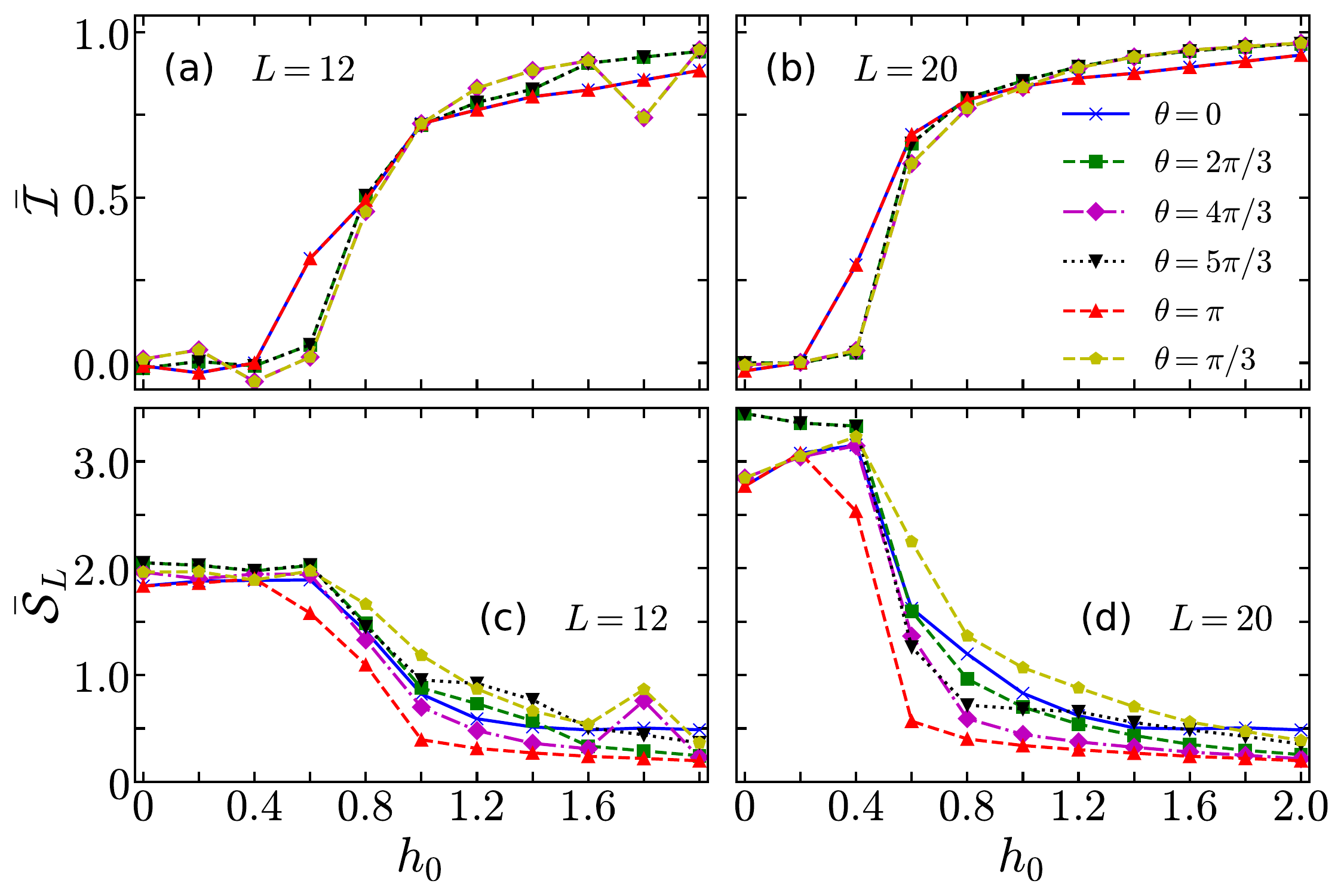}}
	\caption{Localization in as spin $1/2$ chain, for different values of $h_0$ and $\theta$ with $D = 0$, using the average of the imbalance $\langle{\cal I}\rangle$ vs $h_0$ and the average entanglement entropy $\langle S_\ell \rangle$ vs $h_0$. The system is configured in the space $S_z = 0$ with $L = 12$ and $L=20$ with a magnetization island of $1/2$ in the middle of the chain and $-1/2$ at the ends.}
	\label{fig1}
\end{figure}
\subsection{Spin-$1/2$ system: effect of magnetic field}
For the sake of completeness, we reproduce and elaborate on the main results obtained by one of us in Ref.~\cite{Vernek2022} for spin-1/2 SMBL. We validate that the fundamental characteristics are qualitatively well captured at a scale involving $12$ and $20$ spins.
In Fig.~\ref{fig1} we use imbalance and entanglement entropy to show how the system evolves from a thermalized to  localized regime as $h_0$ increases. We use an initial state within  the Hilbert subspace  with $S_{\rm tot}^z=0$ consisting of an island of spin \emph{up} at the center of the chain. For a given value of $h_0$ we perform time evolution up to $J_0t=500$ just as in Ref.~\cite{Vernek2022}. Since both ${\cal I}(t)$ and $S_\ell(t)$ oscillates around a fixed value for large $t$, we take their respective  time average, $\bar {\cal I}$ and $\bar S_\ell$, for $J_0t\in [400,500]$.

In Figs.~\ref{fig1}(a) and \ref{fig1}(c), $\bar {\cal I}$ as a function of $h_)$ is depicted for system sizes $L=12$ and $L=20$, respectively, for various value of $\theta$. We observe that for   $h_0=0$ $\bar {\cal I} \approx 0$ but increases with increasing $h_0$ reaching almost the unity for $h_0=2$ for all value of  $\theta$. The vanishing  $\bar {\cal I}$ for small $h_0$ reveals that the local information of the initial state is lost at long time though the unitary dynamics, while large value of $\bar{\cal I}$ indicates that the information of the initial state is kept local at long times. There is an interesting qualitative agreement between the results for $L=12$ and $L=20$, which shows that localization can be observed for relatively small chains.  Similarly the entanglement entropy shown in  Figs.~\ref{fig1}(c) and \ref{fig1}(d) (for $L=12$ and $L=20$, respectively) confirms  localization in the system as $h_0$ increases. Since the initial state is a product state, with entanglement entropy $S_{L}(t)=0$, large values of $\bar S_L$ for $h_0\rightarrow 0$ indicate that the system becomes entangled, signifying the spread of information throughout the entire system in the thermalized regime. Conversely, vanishing $\bar S_L$ in the localized regime for large $h_0$ confirms that information is kept local at long times. It is noteworthy to mention a symmetry pertaining to the imbalances $\theta$ and $\theta+\pi$, as previously highlighted in  Ref.~\cite{Vernek2022}.

In what follows we will turn our attention to systems with $S>1/2$ on which  $D$ exerts non-trivial effects. The natural progression leads us to the case of $S=1$. However, in this instance, flipping a given spin from $S_z=-1$ to $S_z=+1$ requires passing through  $S_z=0$,  which remains inert to the magnetic field along the $z$-direction.  To explore the interplay between $h_0$ and $D$, the case of $S=1$ does not encompass the most general scenario. Consequently, we shift our focus to the case of $S=3/2$, which presents a richer set of dynamics. Further details regarding the $S=1$ case can be found in Appendix \ref{appendix_A}.

\subsection{Spin-$3/2$ system: effect of magnetic field and single-ion anisotropy} 

\begin{figure*}[!htbp]
	\centering
	\subfigure{\includegraphics[clip,width=3.56in]{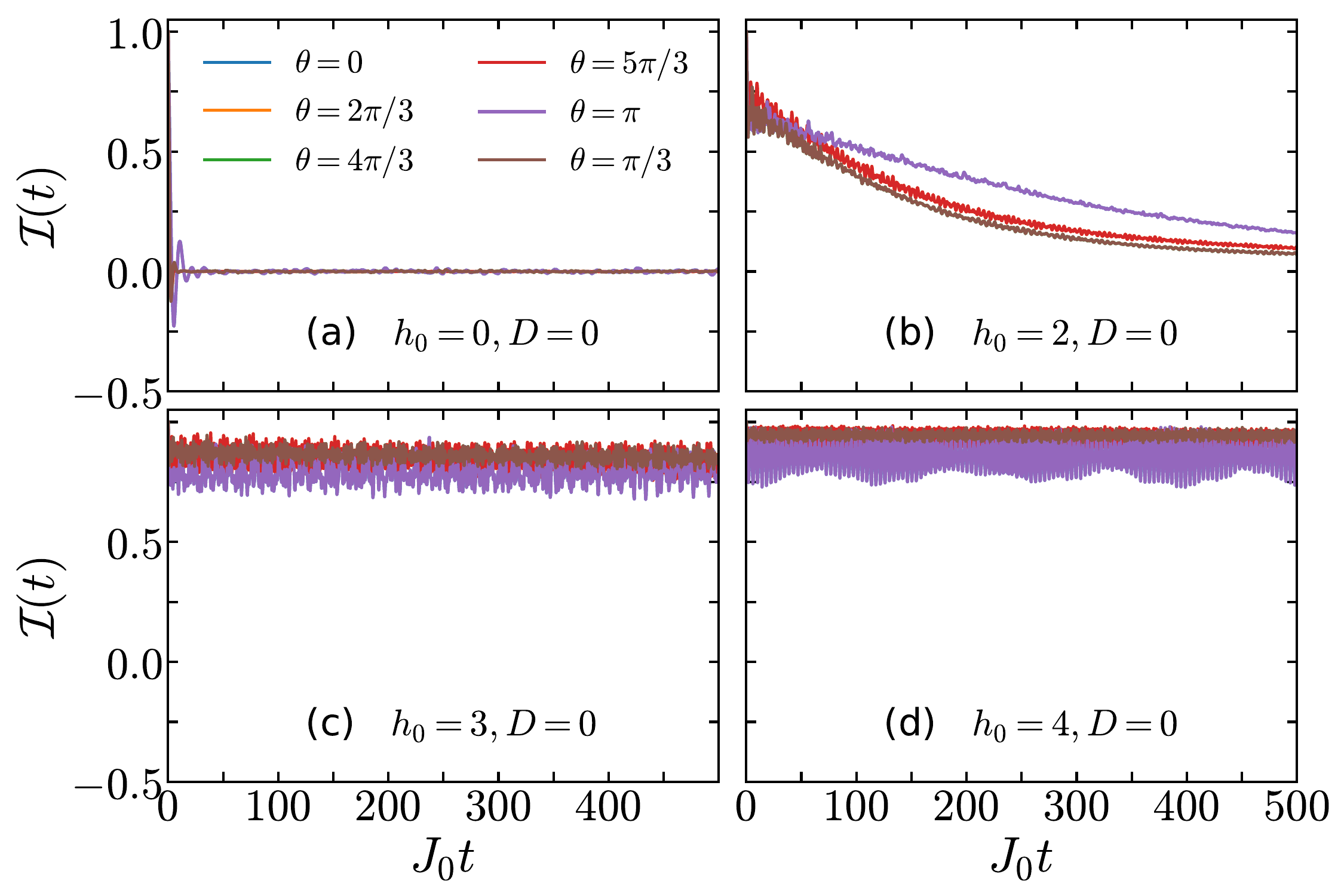}
        \includegraphics[clip,width=3.56in]{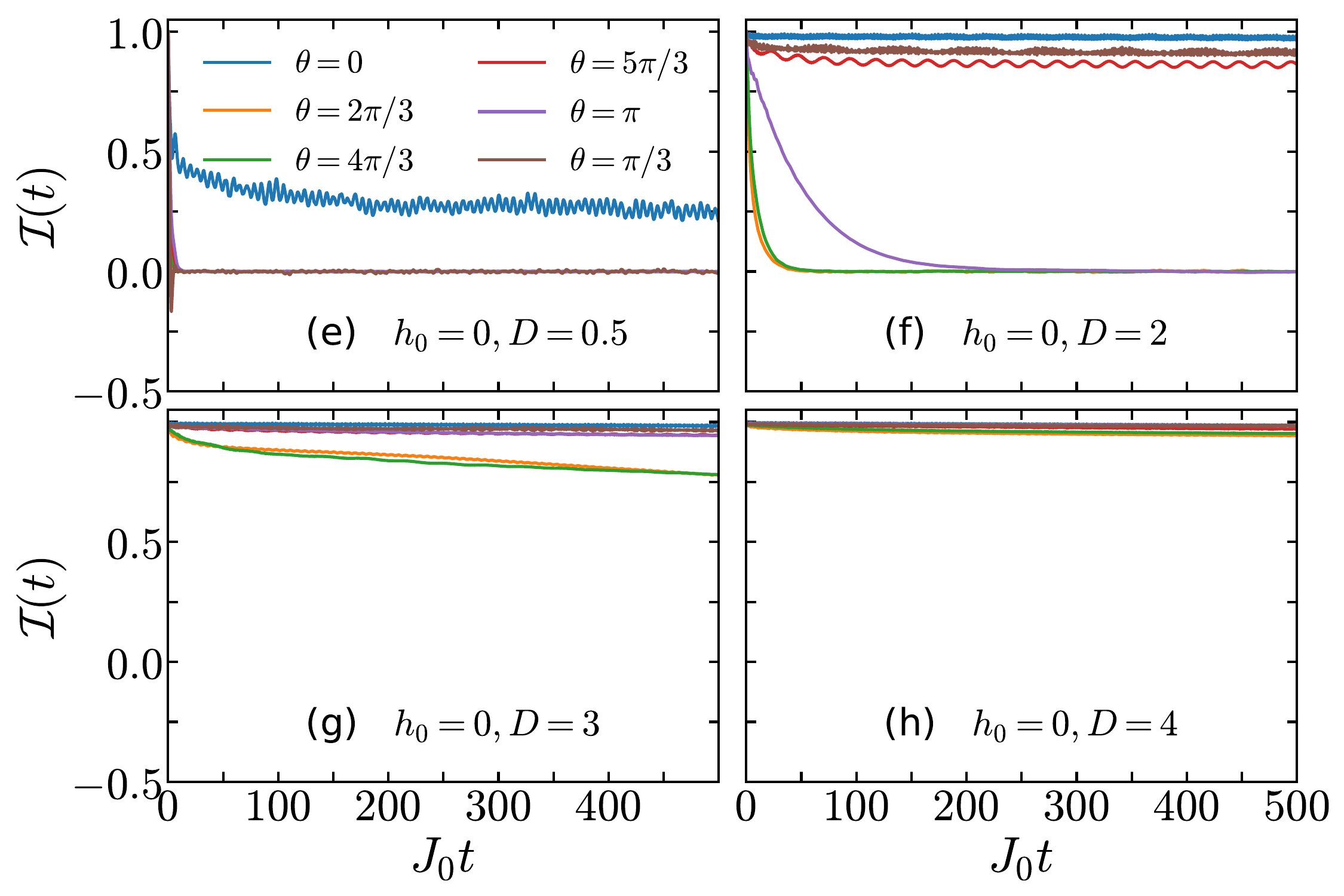}}
	\caption{Imbalance as a function of time for the stark many-body localization (for increasing $h_0$ and $D=0$) [panels (a)-(d)] and for $h_0=0$ and increasind $D$ [panels (f)-(h)].  Different curves corresponds to different values of  $\theta$ (see legend). The system consists of $L=12$ spins and the initial state corresponds to an island of of spins with $S^z_j=3/2$ in the middle of the chain, while all other has $S^z_j=-3/2$.} 
	\label{fig2}
\end{figure*}

The $S=3/2$ case offers a much richer Hilbert space as compared to $S=1/2$. In this case each spins exhibit $S^z = (3/2,1/2,-1/2,-3/2)$ where the two \emph{local} subspaces defined by $|S_j^z|=1/2$ and $|S_j^z|=3/2$ are both active to external magnetic field gradient $h_0$ and anisotropy $D$. This leads to more complex dynamics of the system towards thermalization. We will first study the localization in the presence of $h_0$ and contrast with the case of $S=1/2$ and then investigate the effect of $D$ alone. In a third stage we will study the interplay between $h_0$ and $D$ in the dynamics of the system.

Following the same strategy used before, we will analyze the time-evolution for an initial state where there is an island of $L/2$ spins with $S_j^z=-3/2$ centered at $L/2$ while the rest of the system has $S_j^z=3/2$. Like the previous initial state we used for $S=1/2$, this  is a non-equilibrium product state, hence non-entangled. The effect of $h_0$ in the time evolution of this state is shown in Fig.~\ref{fig2}(a)-\ref{fig2}(d) for $h_0=0$, $h_0=2.0$, $h_0=3.0$  and $h_0=4.0$, while keeping $D=0$. In Fig.~\ref{fig2}(a), corresponding to the case $h_0 = 0$, the Hamiltonian of the system is exclusively governed by the Majumdar-Ghosh exchange terms. We observe rapid thermalization in the system, leading to the decay of ${\cal I}(t)$. When the magnetic field gradient is increased to $h_0 = 2$, as depicted in Fig.~\ref{fig2}(b), the system continues to undergo thermalization for all angles, albeit at a slower pace for every $\theta$. Notably, ${\cal I}(t,\theta)={\cal I}(t,\theta+\pi)$,  a symmetry previously noted for spins $S=1/2$. Fig.~\ref{fig2}(c) and \ref{fig2}(d) exhibits the results  $h_0=3$ and $h_0=4$, respectively. In both situations, where the imbalances remain close to unity, indicating localized states, the observed persistence of these imbalances in systems with spin $3/2$ strongly suggests the occurrence of SMBL in lattices with higher-spin configurations

We now turn our attention to the impact of single-ion anisotropy on the system dynamics, utilizing the same initial states as in the previous analysis. In Fig.~\ref{fig2}(e)-\ref{fig2}(h) we set $h_0=0$ and present the dynamics of ${\cal I}(t)$ for $D>0$. Specifically, Fig.~\ref{fig2}(e) displays the results for $D=0.5$ across various values of $\theta$. All the curves exhibit rapid thermalization, except in the specific case of $\theta=0$, where the imbalance remains finite for an extended duration, up to  $J_0t=500$. This observation strongly indicates that anisotropy has the potential to induce localization, with a notable dependence on the angle $\theta$. Indeed, for $D=4$, as shown in Fig.~\ref{fig2}(h), the largest $D$ considered here, the system remains localized, as all imbalances remain close to unity. The localization observed for large $D$ can be easily understood by observing that for $D\gg J$, the initial state becomes \emph{almost} an eigenstate of the system, apart from some very small perturbation introduced by $J_1$ and $J_2$. In this scenario, localization is trivially achieved. The intriguing situation arises in the competing regime where $D\sim J_0$. Let's take a closer look at Figs.~\ref{fig2}(f) (where $D=2$) in more detail. Observe that the curves for $\theta=0$, $\theta=\pi/3$ and  $\theta=5\pi/3$ remains close to unity, indicating strong localization. Conversely, for $\theta=2\pi/3$, $\theta=\pi$ and $\theta=4\pi/3$ the system undergoes thermalization. Note that $J_2$ is the same for both $\theta=0$ and $\theta=\pi$, yet the dynamics is entirely distinct. This difference can be attributed solely to the sign of $J_1$. In other words, thermalization is primarily governed by nearest-neighbor spin-flip processes facilitated by $J_1$. To comprehend the mechanisms leading to localization (thermalization) $J>0$ ($J<0$) at large $t$, let us analyze the early stages of the demolishing of the domain wall at a given interface between spins $S^z=+3/2$ and $S^z=-3/2$.  The very first process that  destroys the interface is given by the term $JS_i^-S_{i+1}^+$ acting on the interface $\ket{\cdots 3,3,\bar 3,\bar 3\cdots}
$, resulting in an smoother interface  $\ket{\cdots 3,1,\bar 1,\bar 3\cdots}$. 
Due to finite $D$, there is a cost associated with smoothing the domain wall, as the local energy in sites with $S_j^z = \pm 1/2$ significantly differs from that in sites with $S_j^z = \pm 3/2$. This energy difference acts as a barrier that can only be surpassed through virtual processes during the system's time evolution.

\begin{figure}[h!]
\centering
\subfigure{\includegraphics[clip,width=3.45in]{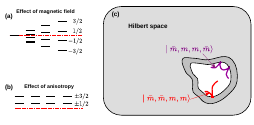}}  
\caption{Effect of a linearly varying magnetic field (a) and a constant anisotropy (b) on the local energy of a spin-$S=3/2$ chain. (c) Representation of the Hilbert space and typical initial state within the  Hilbert space for a  spin-$3/2$ chain. White regions encloses only states in which all spins 
 have their spin projection on the $z$-axis is maximum or minimum, i. e., their projections are $S_j^z\pm 3/2$. Darker region contains states in which some of their spins has projections $S^z_j=\pm 1/2$. The  red and violet curves represent an initial states with single and double domains wall, respectively. The  dots marks the active regions upon time $t=0^+$ in the time evolution operator $\exp(-iH 0^+/\hbar)$.
 }
\label{fig3}
\end{figure}
%


To gain a deeper understanding of the physical origin of this localization, one can explore the limit of very large $|D|$ or, equivalently, $J\rightarrow 0$. For simplicity, let us assume $D>0$. When $J=0$ our initial state is an eigenstate of the Hamiltonian of highest energy, since all spins have the maximum $|S^{z}_j|$.  In fact, any product state constructed from the $S^z$ basis is an eigenstate of the Hamiltonian for $J=0$. The spectrum of the Hamiltonian consists of a series of $delta$-peaks among which the initial state corresponds to one of the states of highest energy (for $D>0$). For the sake of clarity, in Figs.~\ref{fig3}(a) and \ref{fig3}(b) we illustrate the energy cost associated with placing a spin with different $S_j^z$ along the chain, considering a constant-gradient magnetic field $h_0$ and a anisotropy $D$, respectively. Figure \ref{fig3}(c) represents the entire Hilbert space of the system, with the white region denoting the sector of the Hilbert space where the states have spins fully aligned along the $z$-direction. Consider the simple case of the initial state $\ket{\cdots \bar 3,\bar 3,\bar 3, 3,3,3 \cdots}.$ This state exhibits a single domain wall, represented by the red line within the sector of the Hilbert space indicated by the white region. The point where the curve touches the border of the white region represents the domains wall. At $t=0^+$, only the region represented by the red line is active. The purple line indicates a state with a double wall, featuring two active regions. In the vicinity of the white sector, there are states where the two spins around the wall have projections $\pm 1/2$, such as the state $\ket{\cdots \bar 3,\bar 3,\bar 1, 1,3,3 \cdots}$. One can readily confirm that for $J=0$, this state is trivially localized, but for finite $J=0$, its temporal evolution becomes highly complex. Consider that at $t=0$, the system is prepared in the initial state with a single domain wall. At $t=0^+$ the only active region of the state is around the domain wall, since $H$ acting on this state   modifies only the spins at the vicinity of the domain wall. Now, upon time evolution, in order to reach a thermal situation, the mechanism involves states with  $S_j^z\pm 1/2$ in the vicinity of the wall (represented by the dark gray region). It turns out that, for $D>0$ theses intermediate states has of much different energies for $D \gg J$. Therefore, for a given $J$ fixed, by increasing $D$ these sector becomes isolated from the rest of the Hilbert space.

\begin{figure}[h!]
	\centering
	\subfigure{\includegraphics[clip,width=3.40in]{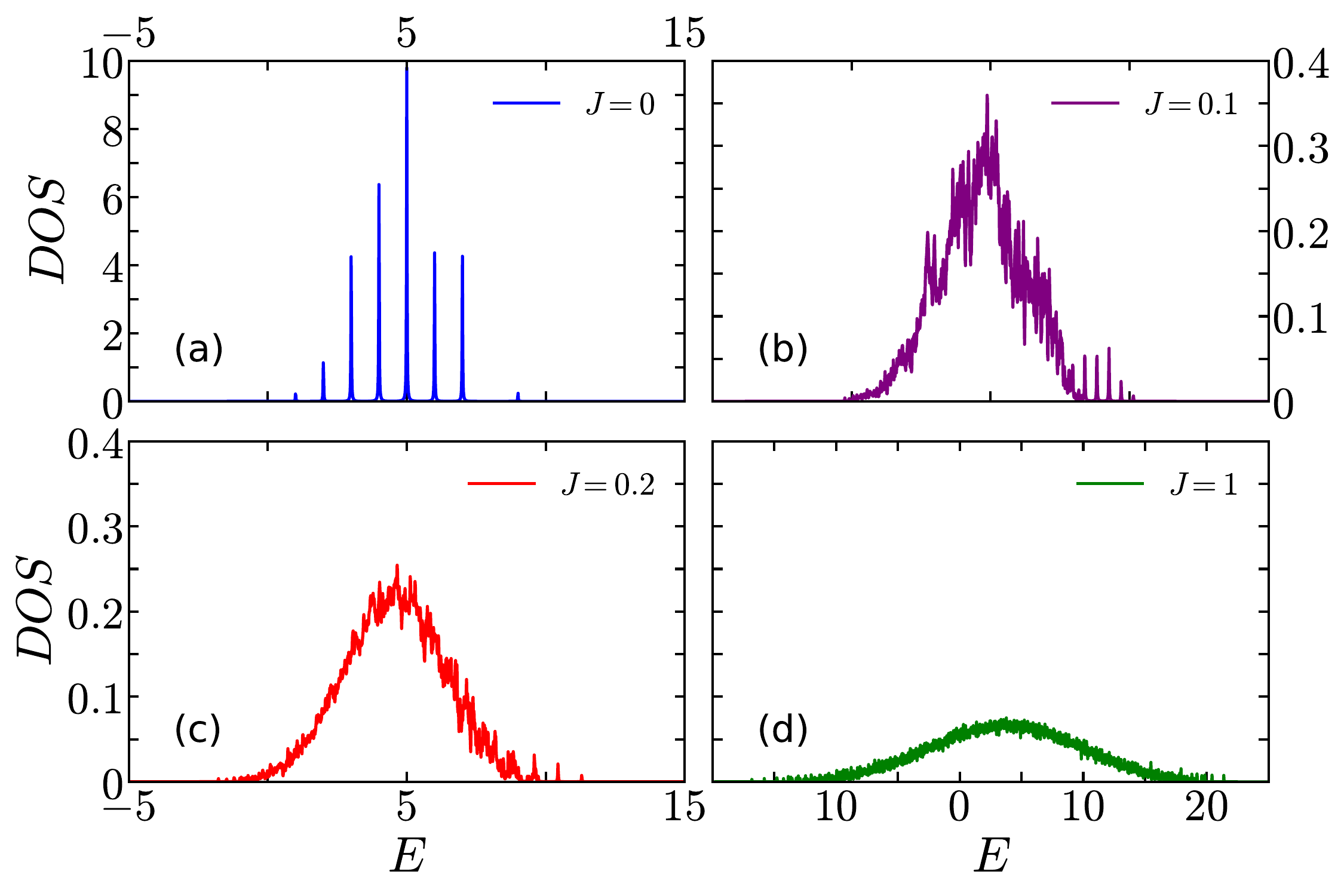}}
	\caption{Density of states as a function of energy for $D=0.5$ and $\theta=0$ for $J=0$ (a), $J=0.1$ (b), $J=1$ (c) and $J=0.4$ (d). For all of them, we use the subspace of $S_z=0$, with the spin-up magnetization island in the middle for L = 8.}
	\label{fig4}
\end{figure}

In Figure \ref{fig4}(a)-\ref{fig4}(d), we present the density of states for a system of $L=8$ spins $S=3/2$ for various values of $J$ and fixed $D=0.5$. In the absence of $J$($J=0$) in \ref{fig4}(a), the initial state precisely corresponds to one of the eigenstates associated with the rightmost degenerate energy peak.  As $J$ increases, these peaks hybridize with other states, resulting in peak splitting. For instance, at 
$J=0.1$, the peaks are split due the the hybridization with the rest of the band. Therefore, the effect of  $D>0$ is evident in pushing the initial state to higher energy, creating a \textit{"gap"} between this state and the rest. Consequently, the entire white region of the Hilbert space, as illustrated in \ref{fig3}(c), becomes isolated from the rest. In contrast to QMBS, where the Hilbert space fragmentation is driven by emergent conserved quantities, the observed suppression of thermalization in this system arises from the inhibition of spin flip processes due to energy constraints.
The picture drown above is useful to understand  why for a moderate value of $D$, such as $D=2$ and $J=1$ shown in Fig.~\ref{fig2}(f),   by changing $\theta=0$ to $\theta=\pi$ (or equivalently $J$ to $-J$), the system goes from localized to thermalized. The reason is that, spectrum is not symmetric when $J\rightarrow -J$. In fact, for $J>0$ the spectrum extends more towards negative side of the energy axis. The opposite occurs   $J<0$. Therefore, for a $D$ fixed, the energy ``gap'' between the sector of the initial state is larger for $J>0$, thus localization is obtained more easily. The evidence that increasing $D$ suppresses dynamics of spins around the domain walls, places the localization mechanism here in the same perspective of the SMBL. As discussed in Ref.~\cite{PhysRevB.105.L140201}, in SMBL, only for infinite $h_0$ the spins a completely frozen by ``local energy'' constraint. Likewise, here, the spin dynamics will be fully frozen only at infinite $D$. Nevertheless, our results suggest localization already for $D \sim 2J$ as in Fig.~\ref{fig2}(b).

\begin{figure}[!htbp]
\centering
\subfigure{\includegraphics[clip,width=3.40in]{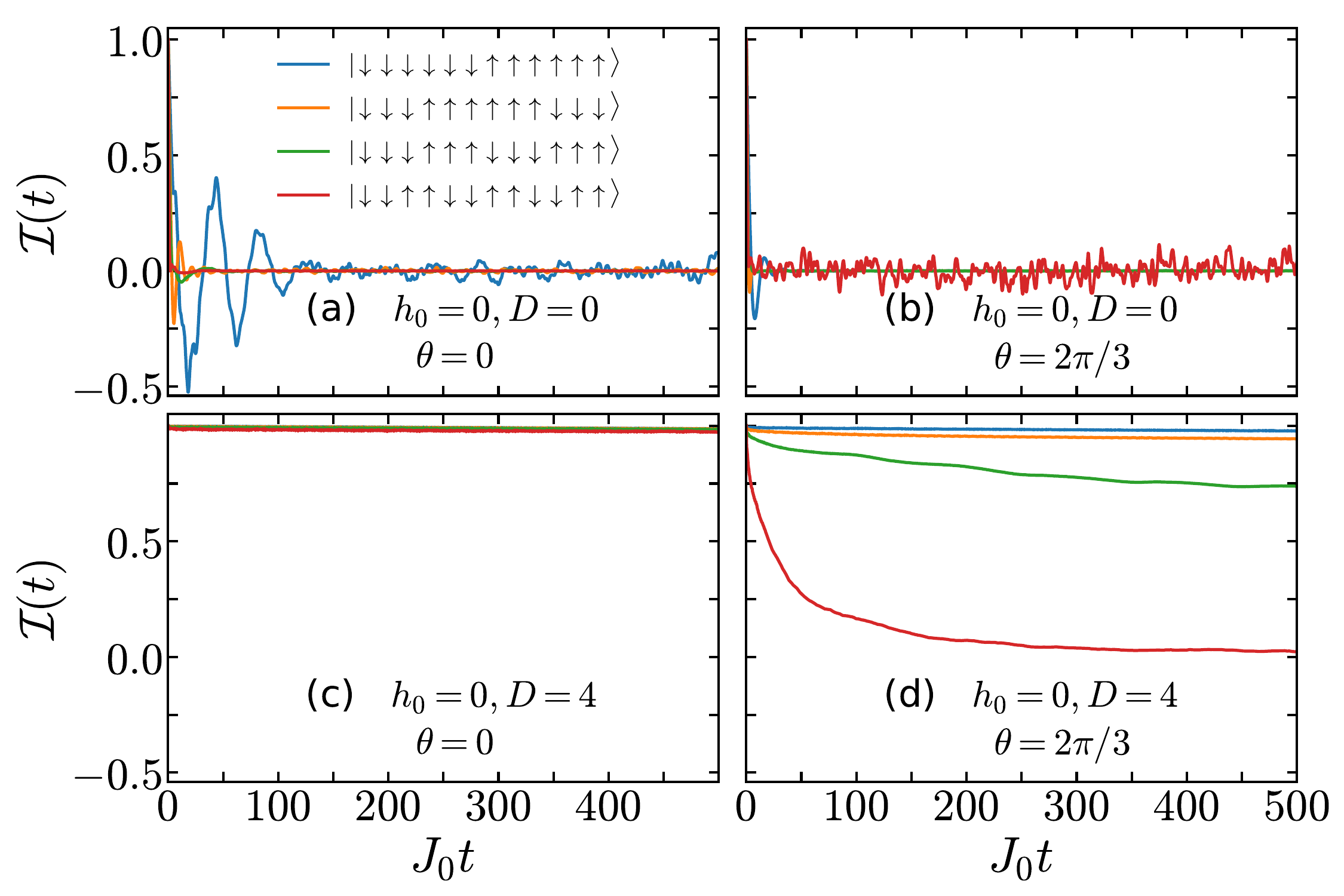}}
\caption{Time evolution of imbalance for  different initial conditions (different curves). Upper and lower panels correspond to $D=0$ and $D=4$, while left and right panels correspond to $\theta=0$ and $\theta=2\pi/3$. For all panels  use a chain of spin $3/2$ and $L = 12$ and $h_0=0$. All initial states here lies on the $S_z=0$ subspace.
} 
\label{fig5}
\end{figure}

Since the demolishing of domain walls of the initial states occurs because of the active regions around the interfaces, let us analyse how the number of walls modifies the localization/thermalization.  In Fig.~\ref{fig5} shows the time evolution of the imbalance  starting at an initial states containing one, two, three and five domains walls, all  states within the zero total magnetization. Again, we use a chain of $L=12$ spins $S=3/2$ and set $h_0=0$. Panels \ref{fig5}(a) and Fig.~\ref{fig5}(b) show the case of $D=0$ and $\theta=0$ and $\theta=2\pi/3$. For this we observe thermalization for all initial states shown. Now, for $D=4$ (lower panels), while for $\theta=0$ the system is still localized, as shown in Fig.~\ref{fig5}(c), for $\theta=2\pi/3$ localization becomes poorer as the number of walls increases. For instance, for $\theta=2\pi/3$, note that for five interfaces the system thermalizes, while remains localized for one and two walls. This is because, the number of active regions increases with the number of domain walls as illustrated in Fig.~\ref{fig3}(a), thus favoring thermalization. This shows  that the larger is the number of interfaces, the larger is the value of $D$ and $h_0$ necessary to produce the same localization, similar to what was previously reported in Refs~\cite{Doggen2021} for SMBL.

\subsection{Interplay between anisotropy and magnetic field}
\begin{figure*}[!htbp]
\centering
\subfigure{\includegraphics[clip,width=3.56in]{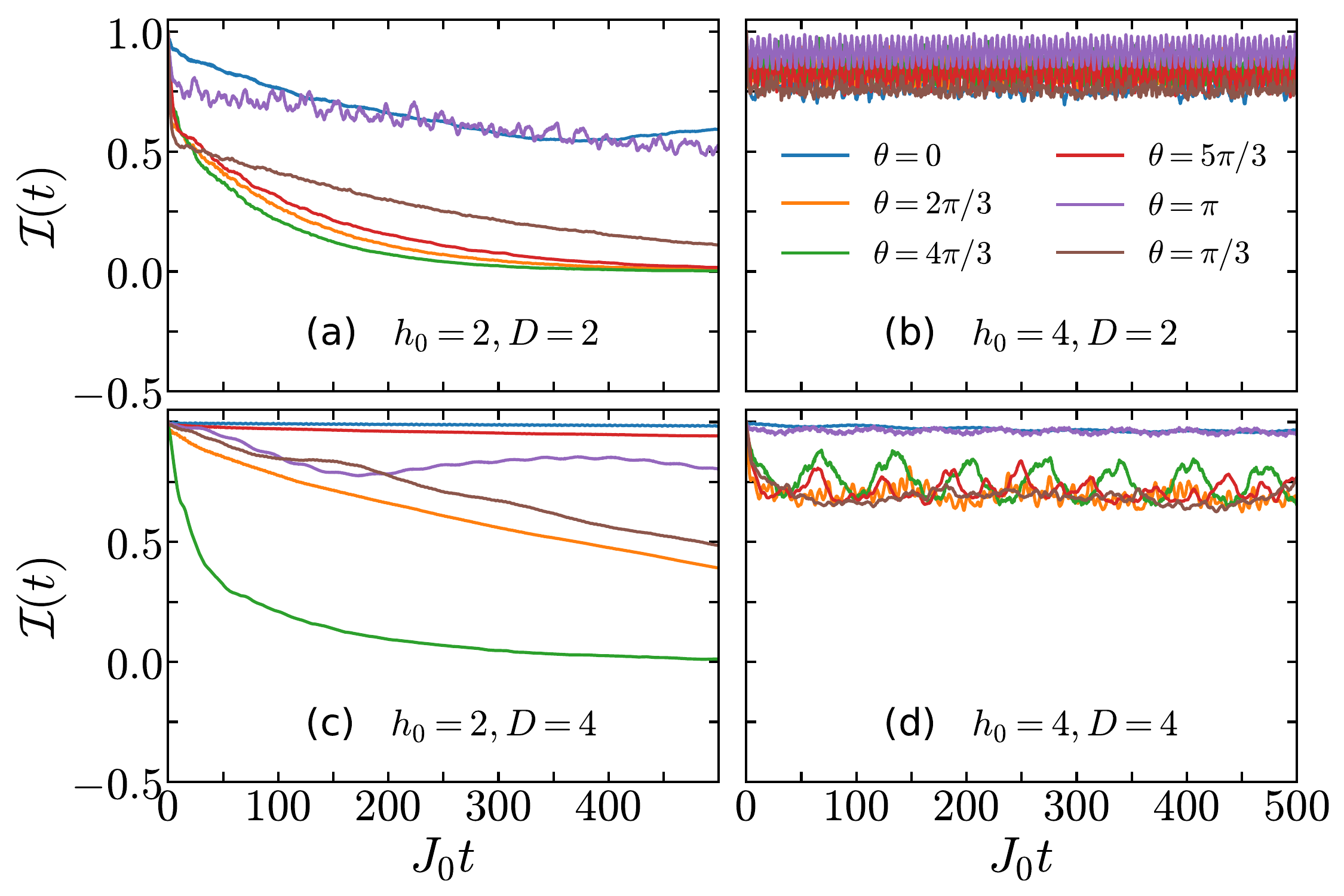}
\includegraphics[clip,width=3.56in]{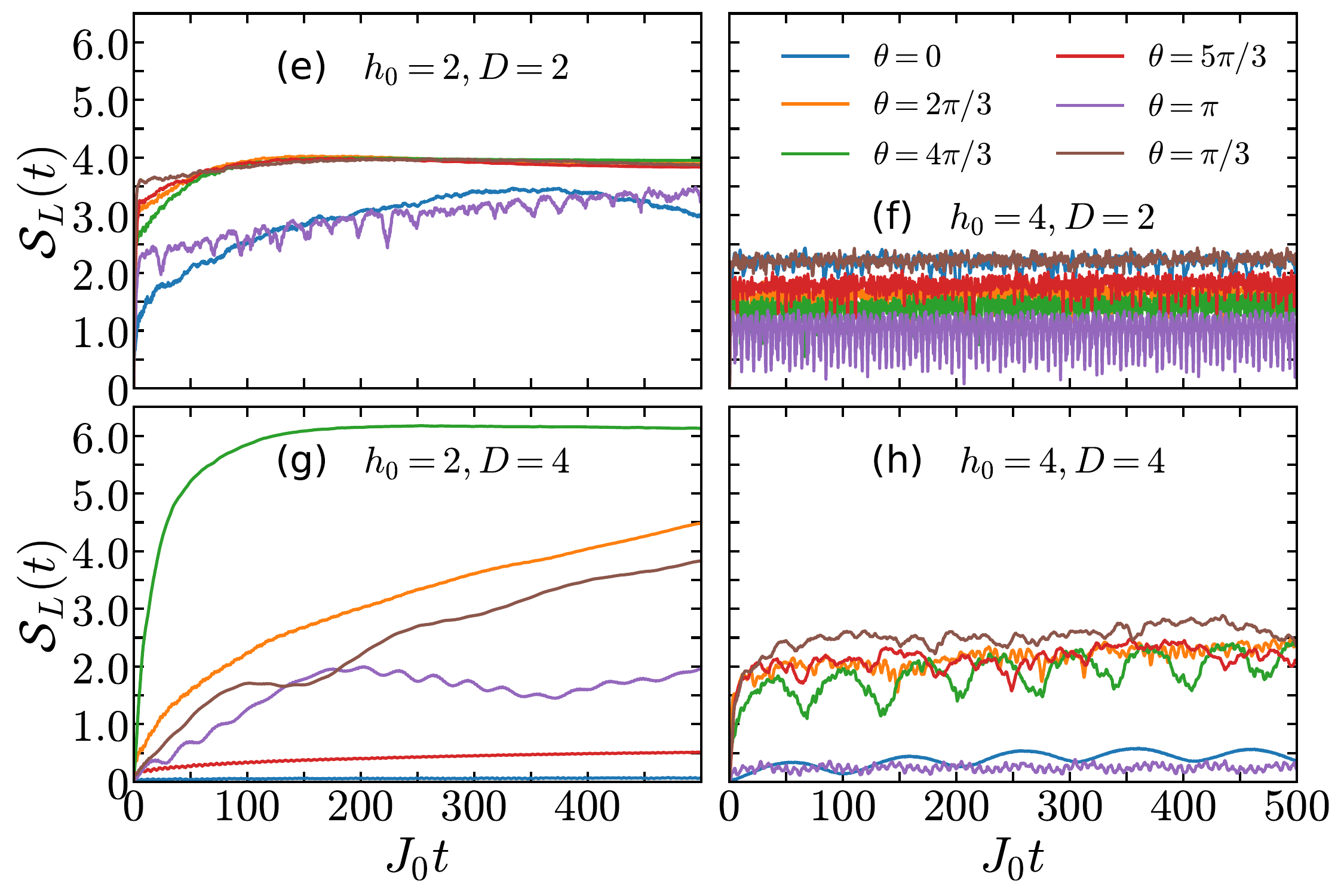}}
\caption{Time evolution of imbalance (a)-(d)  and entropy (e)-(h) for different values of $h_0$ and $D$. The system and initial condition here is the same as the one used in Fig.~\ref{fig2}. The curves corresponds to different values of 
$\theta$ (see legend). } 
	\label{fig6}
\end{figure*}

\begin{figure}[!htbp]
    \centering
    \subfigure{\hspace{0.09cm}\includegraphics[clip,trim=0cm 0cm 0cm 0cm,width=2.75in]{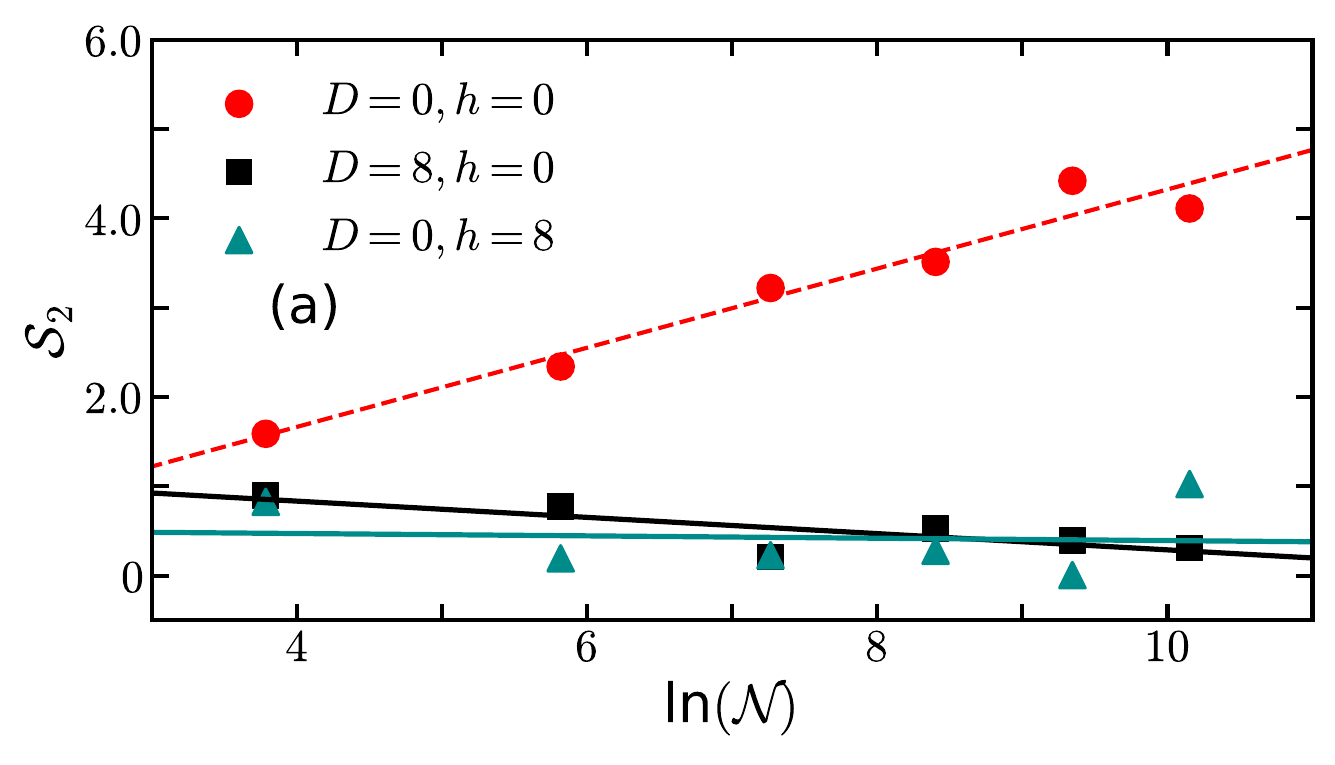}}
    \vskip-0.35cm
    \subfigure{\includegraphics[clip,width=2.78in]{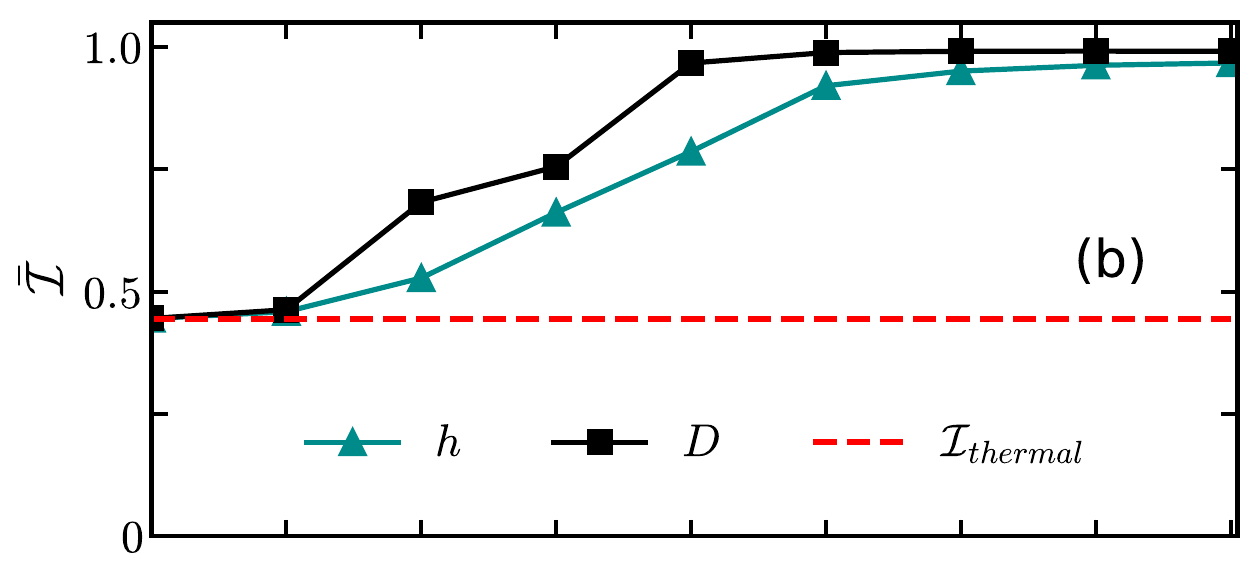}}
    \vskip-0.42cm
    \subfigure{\includegraphics[clip,width=2.78in]{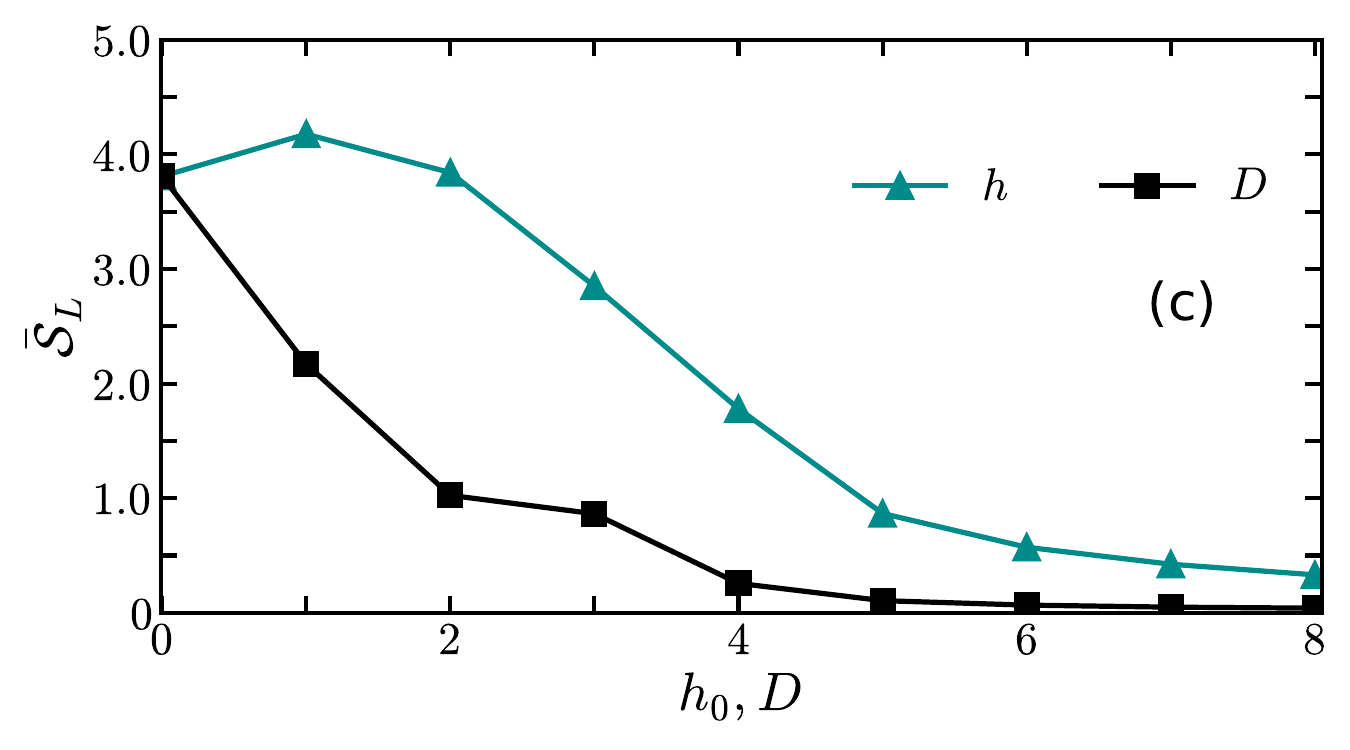}}
    \caption{(a) Participation entropy vs $\ln {\cal N}$ for $(D,h_0)=(0,0)$
    (red circles),  $(D,h_0)=(8,0)$ (black squares) and $(D,h_0)=(0,8)$ (cyan triangles). Dashed lines corresponds to a linear fitting of the curces and serves to guide the eyes.  To change the size of the Hilbert space,  $L$ includes all even numbers form $4$ to $14$. (b)  and (c) shows, respectively, latter time average of imbalance and entropy  as function of $D$ (for $h_0=0$) (red squares) and  $h_0$ (for $D_0=0$) (blue triangles) for $L=12$. For all panels, we set $\theta=0$ ans use an initial state of type $\ket{\cdots,\bar 3,3,3,\bar 3,\cdots}$, which corresponds to an island of spins $S_j^z=3/2$ in the middle of a chain of spins $S_j^z=-3/2$.}
    \label{fig7}
\end{figure}
So far we have analysed separately the SMBL, induce by the magnetic field gradient, and the localization induced by anisotropy. Let us now address the interplay between these two terms of the Hamiltonian.  In Fig.~\ref{fig6}(a)-\ref{fig6}(d) we repeat the same calculation of Fig.~\ref{fig2}, but now for both $h_0$ and $D$ finite. Figure~\ref{fig6}(a) shows the imbalance for $h_0=D=2$. We observe that while localization were observed for some values of $\theta$ for $D=2$ and $h_0=0$ shown in Fig.\ref{fig2}(f), now localization is no longer observed for any value of $\theta$. Interestingly, if we make $h_0=2D=4$, as shown in Fig.~\ref{fig6}(b), localization is again recovered. However, for $D=2h_0=4$, localization is observed  only for some  values of $\theta$, in fact for those closer to $\theta=0$ of $\theta=2\pi$. Now, for $\D=h_0=4$, all the curves  approach to unity at large $t$. Despite the complexity of the behavior of the various curves, results indicate that for moderate values of $D\approx h_0$, they compete against each other and fail to localize the system. However, if both are much larger than $J$ localization can be recovered again. The competition between $D$ and $h_0$ observed here can be understood as follows: the energy price that strongly suppresses spin dynamics around the domain wall for $D$ finite (and $h_0=0$) is now, to some extent, compensated by finite $h_0$. Likewise, the suppression of the dynamics for finite $h_0$ (and $D=0$) is now energetically relaxed if $D$ is comparable to $h_0$.

In Fig.~\ref{fig7}(a) we show the participation entropy as function of ${\cal N}$ (the dimension of the Hilbert space for $(D,h_0=0)=(0,0)$ (red circles), $(D,h_0)=(8,0)$ (black squares), and  $(D,h_0)=(0,8)$ (cyan triangles). Here we use an initial state consisting of chains of even number of sites from $L=4$ to $L=14$. The initial states are of the type $\ket{\cdots \bar 3, 3,3, \bar 3 \cdots}$. For all system size, the initial state contains an island of only two spin $S_j^z=3/2$. For $(D,h_0=0)=(0,0)$ we note the $S_2$ increases linearly with $\ln{\cal N}$, as expected for an ergodic regime. In contrast, for  
both cases $(D,h_0)=(0,8)$ and $(D,h_0)=(8,0)$, $S_2$ remains nearly constant. This indicates that for the ergodic regime, the number of eigenstates that participates on the dynamics increases with the dimension of the Hilbert space. In the localized regime, on the other hand, only a limited portion of the eigenstates contributes to the dynamics. 

Figure~\ref{fig7}(b) shows $\bar{\cal I}$ vs $D$ for $h_0=0$ (green squares) and $\bar{\cal I}$ vs $h_0$ for $D=0$. We use a chain of $L=12$ and the same initial state as in Fig.~\ref{fig7}(a) and set $\theta=0$. Again, $\bar{\cal I}$  is the average of ${\cal I}(t)$ for $t \in [400,500]$. We note that both curves evolves quite nicely  from a thermalized ($\bar{\cal I}\approx 0.5$) to a localized ($\bar{\cal I}\approx 1$) regime as $D$ and $h_0$ increases from zero to $8$. Interestingly, the crossover occur for similar values of $D$ and $h_0$. This is corroborated by the entropy $\bar{S}_L$ depicted in Fig~\ref{fig7}(c), where it is evident that ${\bar S}_L$ vanishes if either  $h_0$ and $D$ increase, indicating the localization of the states.

To further elucidate the numerical results, we employ the Holstein-Primakoff (HP)
transformation; mapping spin operators in a lattice system with spin$-S$ to
bosonic operators: $S_{j}^{z}=(S-n_{j})$, $S_{j}^{+}=\sqrt{2S-n_{j}}a_{j}$, and $S_{j}^{-}=a_{j}^{\dagger }\sqrt{2S-n_{j}}a_{j}$, where $a_{j}^{\dagger
}(a_{j})$ is a bosonic creation (annihilation) and the number $n_{j}=a_{j}^{\dagger }a_{j}$ operator at site $j$. Subsequently, a $1/S
$ expansion enables us to derive an effective description of the quantum
dynamics, expressed as $H_{HP}=H_{G}+H_{1}+H_{2}+\mathcal{O(}1/S)$. The term 
$H_{G}$ characterizes the classical ground-state energy, while $H_{1}$
represents the magnon dispersion, both contingent on the parameters $\left\{
J_{1},J\,_{2},D,h_{0}\right\} $, yet not influencing localization. The
leading contribution for comprehending the interplay between $h_{0}$ and $D$ emerges from $H_{2},$ since this term and higher-order ones, describes magnon-mangnon interactions. To simplify, we consider only nearest-neighbor interaction, i.e., $J_{1}=J>0,J_{2}=0$ and $D>0$. The HP representation leads to the effective Hamiltonian
\begin{eqnarray}
H_{\rm HP}&&\approx\sum_{j=1}^{L}\left( 2DS+2JS-h_{j}\right) n_{j}
-JS\sum_{j=1}^{L-1}(a_{j}^{\dagger }a_{j+1}+H.c.)\nonumber \\
&&+\frac{J}{8}%
\sum_{j}\left[\left(a_{j}^{\dagger }\right)^{2}a_{j}a_{j+1}+\left(a_{j+1}^{\dagger }\right)^{2}a_{j+1}a_{j}+H.c.\right]\nonumber \\
&&+D\sum_{j}^{L}n_{j}(n_{j}-1)+\frac{J}{2}\sum_{j=1}n_{j}n_{j+1}+H_{G}
.
\end{eqnarray}

The form of $H_{\rm HP}$ is similar to the generalized Bose-Hubbard model in a
tilted optical lattice, with on-site chemical potential $\mu
_{j}=2DS+2JS-h_{j}$ (see Ref.~\cite{yao2020many}). The single-ion anisotropy $D$ is mapped into a
nonlinear repulsive on-site interaction $U$, symbolizing the additional energy needed to accommodate more than one boson at each location. In the context of a tilted lattice, where $\mu _{j}=jA$, in which $A$ represents the strength of the tilt, the standard Bose-Hubbard model exhibits SMBL~\cite{yao2020many,taylor2020experimental}, closely resembling the behavior observed in our numerical results. Specifically, at large values of $\ U$, high-energy states exhibit localized features at all
values of $A$. This behavior is inherently connected to the segmentation of the Hilbert space into subbands, a characteristic that becomes more pronounced in regions with strong interactions. It leads to the freezing of dynamics on long time scales for initial states above a certain energetic threshold~\cite{carleo2012localization}. Another parallel emerges in the interplay between $U$ and $\mu _{j}$, where an increase in $U$ corresponds to an increase in the critical Stark field $A_{c}$ necessary for the manifestation of the localization phenomenon. Additionally, the symmetry noted in our results, encompassing both positive and negative single-ion anisotropy $D$, is reflected in the bosonic scenario as an exact symmetry between two Hamiltonians with the sign of $U$ changed and the energy ordering of eigenvalues and eigenvectors reversed.

\section{Conclusions} \label{conclusions}
In summary, we have investigated the dynamics of a closed spin chain modeled as a Majumdar-Ghosh Hamiltonian in the presence of finite-gradient magnetic field $h_0$ inducing SMBL. An additional term accounting for single-ion uniaxial anisotropy $D$ was included. Employing exact numerical calculations, we compute various quantities, such as imbalance and entanglement entropy, which are commonly used to probe localization phenomena. Our results show that the known SMBL observed for $S=1/2$ chains is observed for $S>1/2$. Similarly, in the absence of magnetic field, the presence of anisotropy alone is able to induce localization in the system for $S>1/2$ for certain initial product states where the spins have maximum $S_j^z$ projection. Our results suggest that the localization induced by $D$ results from the suppression of spin dynamics across the well defined domain walls of the initial states. This can be understood as a energy barrier imposed by the anisotropy that has to be overcome in order to scramble the spins from an product initial state where all spins have their maximum or minimum $S^z_j$ projections.  Even though the sector of the Hilbert space containing these family of product states are not completely decoupled from the rest under time evolution, they become progressively isolated as $D$ increases. This is somewhat similar to what was pointed out by Zisling et al, in Ref.~\cite{PhysRevB.105.L140201} in the context of SMBL. More interesting, we found that for $D\approx h_0 \gtrsim J$ the magnetic field and anisotropy compete to localized the system, favoring delocalization. We interpret this as an interplay between energy costs payed to melt the domain walls. In the competing regime, the  energetic price charged by $D$ is somehow compensated by the cost imposed by the $h_0$.
While our exact numerical calculations are currently applicable to few-site systems, we contend that these initial findings hold relevance in the broader context of the dynamics of closed many-body quantum systems. Particularly, they contribute to our understanding of the connections between Hilbert space fragmentation and disordered free localized systems.
It is important to highlight that our results do not rule out the possibility, for example, of the gradual disappearance of localized regions in the Hilbert space in large systems or in the long-time limit. Addressing these questions demands further exploration, both experimentally using probes commonly employed in conventional MBL studies and theoretically through investigations that could, for instance, extend the theory of dynamical l-bits ~\cite{gunawardana2022dynamical} to spin lattices with $S\geq 1$. Finally, it is worth noting that the models examined in this study can be experimentally realized using current quantum simulators, including cold atoms, trapped ions, and superconducting qubits.

\begin{acknowledgments}
The authors acknowledge financial support from CAPES, FAPEMIG and CNPq. EV thanks FAPEMIG (Process PPM-00631-17) and CNPq (Process 311366/2021-0). EV also acknowledge support and hospitality from the Zhejiang Normal University, where part of this work was conducted. This work used resources of the ``Centro Nacional de Processamento de Alto Desempenho em São Paulo (CENAPAD-SP).''
\end{acknowledgments}

\appendix
\section{Spin-$1$ chain} 
\label{appendix_A}
\begin{figure}[h]
\centering
\subfigure{\includegraphics[clip,width=3.20in]{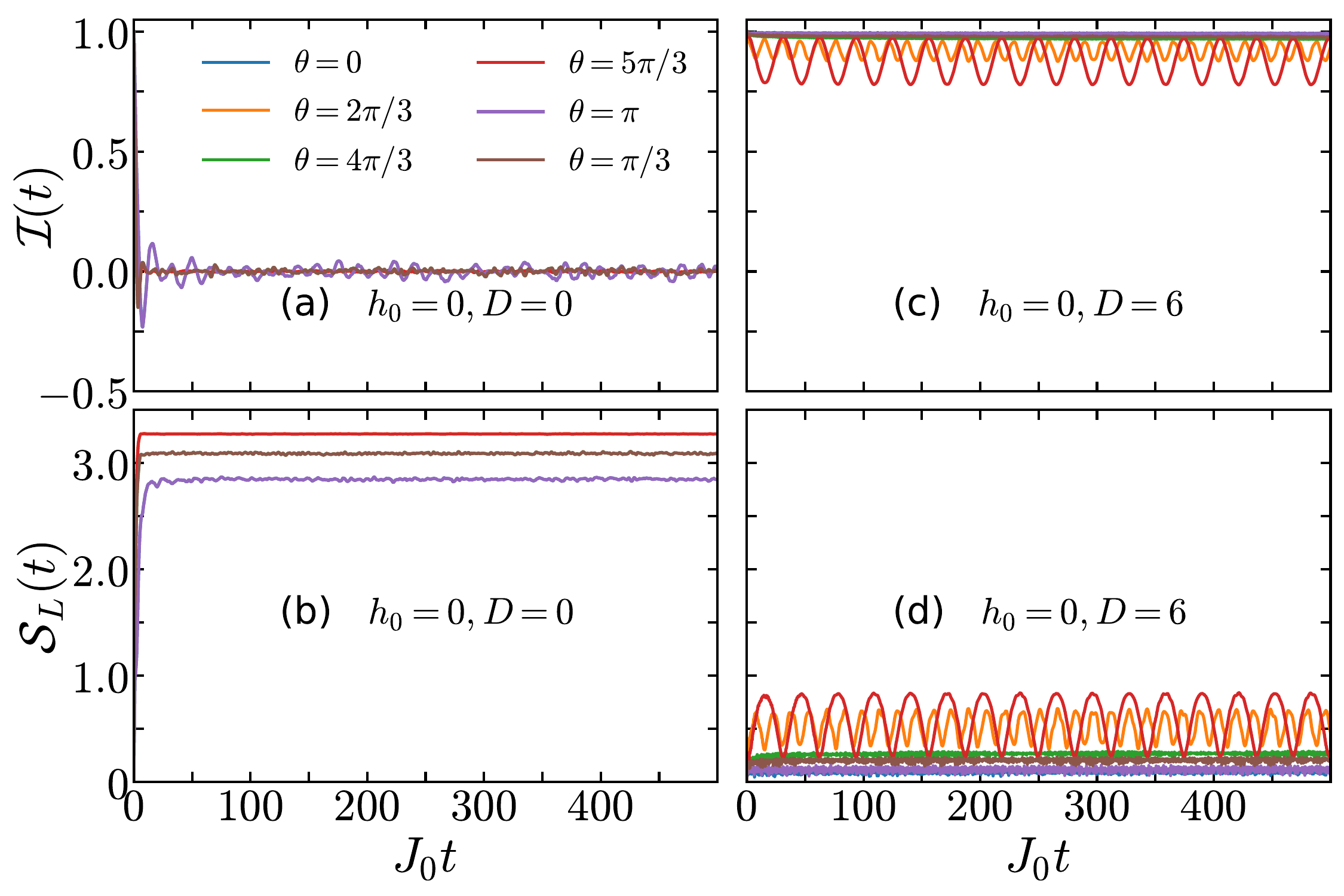}}
\caption{Imbalance (upper) and entanglement entropy (lower) as function of time  for a  spin-$1$ chain for $D=0$ (left) and $D=6$ (right). The curves correspond to different values of $\theta$. We used a chain of $L = 12$ spins with an initial  state consisting of a island of six spins \emph{up} in the middle of the chain, while all other spins are \emph{down} (see text).} 
	\label{fig8}
\end{figure}
Here, we briefly aim to demonstrate that the anisotropy $D$ also prevents thermalization of the $S=1$ spin chain. We consider a chain of $L=12$ and set $h_0=0$. Figure~\ref{fig8}, the upper and lower panels display the imbalance and entropy, respectively. We use an initial state $\ket{\cdots \bar 2,2,2,2,2,2,2 ,\bar 2\cdots}$ and present the results for various angles $\theta$. In Figs.~\ref{fig8}(a) and \ref{fig8}(b) we show ${\cal I}(t)$ and ${\cal S}_L$ for $D=0$. It is noteworthy that the imbalance rapidly drops to zero, while ${\cal S}_L(t)$ indicates the system's tendency to thermalize during time evolution. In Figs.~\ref{fig8}(c) and \ref{fig8}(d), the corresponding results for $D=6$ are presented. In this scenario, we observe that the imbalance remains close to unity, while the entropy consistently stays close to zero. This observation indicates that, in the presence of $D=6$, the system is localized.



\bibliography{references}
\bibliographystyle{apsrev4-2}

\end{document}